\documentclass[%
 ,twocolumn%
 ,secnumarabic%
,amssymb, amsmath,nobibnotes, aps, prl]{revtex4}
\usepackage{graphicx}
\usepackage{docs}%
\usepackage{bm}%
\expandafter\ifx\csname package@font\endcsname\relax\else
 \expandafter\expandafter
 \expandafter\usepackage
 \expandafter\expandafter
 \expandafter{\csname package@font\endcsname}%
\fi

\begin{document}


\title{Spin Filtering in a Nonuniform Quantum Wire with Rashba Spin-Orbit Interaction}

\author{Xianbo Xiao}
\affiliation{Department of Physics, Tongji University, Shanghai 200092, China.\\
College of Science, Jiangxi Agricultural University, Nanchang 330045, China.}
\author{Xiaomao Li}
\affiliation{College of Science, Jiangxi Agricultural University,
Nanchang 330045, China.}
\author{ Yuguang Chen}
 \email{ygchen@mail.tongji.edu.cn}
\affiliation{ Department of Physics, Tongji University, Shanghai
200092, China. }

\date{\today}

\begin{abstract}
We investigate theoretically the spin-polarized electron transport
for a wide-narrow-wide (WNW) quantum wire under the modulation of
Rashba spin-orbit interaction (SOI). The influence of both the
structure of the quantum wire and the interference between different
pairs of subbands on the spin-polarized electron transport is taken
into account simultaneously via the spin-resolved lattice Green
function method. It is found that a very large vertical
spin-polarized current can be generated by the SOI-induced effective
magnetic field at the structure-induced Fano resonance even in the
presence of strong disorder. Furthermore, the magnitude of the spin
polarization can be tuned by both the Rashba SOI strength and
structural parameters. Our results may provide an effective way to
design a spin filter device without containing any magnetic
materials or applying a magnetic field.

\end{abstract}

\maketitle

\section{INTRODUCTION}

Spintronics, aiming to use electron spin instead of charge degree of
freedom to store and communicate information, has drawn considerable
attention in the past decades.$^{1}$ In this field, the Rashba
SOI$^{2,3}$ plays an important role since it has been expected to be
used to manipulate spin states and its strength can be tuned by
external gate voltage conveniently.$^{4-6}$

It is well known that one indispensable requirement of spintronics
applications is to be capable of generating a spin-polarized
current. Various schemes such as spin injection,$^{7}$ spin hall
effect$^{8-9}$ as well as spin filtering$^{10-21}$ have been
proposed to satisfy this goal. Among these schemes, the spin
filtering, which generates a tunable spin-polarized current out of
unpolarized sources, is the most promising candidate. Recently, a
interesting SOI-induced Fano resonance is found in quantum wire with
local Rashba SOC. This effect originates from the SOI-induced bound
state couples to conductance one through Rashba intersubband mixing,
giving rise to pronounced dips in the linear conductance.$^{22}$
These dips at the SOI-induced Fano resonances have the typical Fano
line shape. Apart from the SOI-induced Fano resonance, there is
another type of Fano resonance, that is, structure-induced Fano
resonance, which results from the Fano-type interference between the
continuous subband states and a bound state formed in the system. A
transverse spin-polarized current can be created at these
SOI-induced and structure-induced Fano resonances from a unpolarized
injection has been predicted by Zhai and Xu.$^{17}$ However, the
maximum value of this transverse spin polarization is not more than
0.5 because only at least two propagating modes involved in the
transport can the spin polarized current be created. Thus it is not
known whether a large spin-polarized current can be generated at
other directions at the structure-induced Fano resonance from a
spin-unpolarized injection and whether the magnitude of the spin
polarization can be enhanced by designing an appropriate structure
of the considered system.

In the present work, we calculate the spin polarized electron
conductance for a WNW wire with Rashba SOI by the spin-resolved
lattice Green function method. It is shown that a very large
vertical spin-polarized current can be generated in this system
because of the mirror asymmetry with respect to the longitudinal
axis and the structure-induced Fano resonance. The degree of the
generated vertical spin-polarized current can be tuned by both the
strength of Rashba SOI and the structural parameters. The more
important result is that this vertical spin-polarized current with
strong a robustness against disorder. The organization of this paper
is as follows. In section II, the theoretical model and the
calculation method are presented. In section III, the numerical
results are illustrated and discussed. A conclusion is given in
section IV.

\section{MODEL AND ANALYSIS}

The investigated system is depicted in the inset of Fig. 1(a)
schematically, where a two-dimensional electron gas (2DEG) in the
$(x,y)$ plane is restricted to a WNW quantum wire by a hard-wall
transverse confining potential $V(x,y)$. The 2DEG is confined in a
asymmetric quantum well, where the SOI is assumed to arise
dominantly from the Rashba mechanism. The narrow part of the quantum
wire has a width $W_2$ and a length $L_2$. The wide parts of the
quantum wire have the same width $W_1$ but different lengths $L_1$
and $L_3$, connected to two ideal semi-infinite leads with the same
width at each end. Since we are only concerned with spin-unpolarized
injection, the two connecting leads are nonmagnetic and have a
vanishing SOI. After discreting procedure, a type of tight-binding
Hamiltonian including the Rashba SOI on a square lattice is
obtained,
\begin{eqnarray}
H=\sum\limits_{lm\sigma}\varepsilon_{lm\sigma}C_{lm\sigma}^{\dag}C_{lm\sigma}-t\sum\limits_{lm\sigma}\{C_{l+1,m\sigma}^{\dag}C_{lm\sigma}\nonumber\\
+C_{l,m+1,\sigma}^{\dag}C_{lm\sigma}+H.c\}-t_{so}\sum\limits_{lm\sigma\sigma'}\{C_{l+1,m,\sigma'}^{\dag}\nonumber\\
(i\sigma_{y})_{\sigma\sigma'}C_{lm\sigma}-C_{l,m+1,\sigma'}^{\dag}(i\sigma_{x})_{\sigma\sigma'}C_{lm\sigma}+H.c\},
\end{eqnarray}
where $C_{lm\sigma}^{\dag}(C_{lm\sigma})$ is the creation
(annihilation) operator of electron at site $(lm)$ with spin
$\sigma$. The on-site energy is $\varepsilon_{lm\sigma}=4t$ with the
hopping energy $t=\hbar^{2}/2m^{\ast}a^{2}$, where $m^{\ast}$ and
$a$ are the effective mass of electron and lattice constant,
respectively. The Rashba SOI strength is $t_{so}=\alpha/2a$ and
$\sigma_{x(y)}$ is Pauli matrix.

In the ballistic transport, the spin-resolved conductance is given
by Landauer-B$\ddot{u}$ttiker$^{23}$ formalism with the help of the
nonequilibrium Green function formalism.$^{24}$ The two-terminal
spin-resolved conductance is given by
$G^{\sigma'\sigma}=e^2/hTr[\Gamma_{L}^{\sigma}G_{r}^{\sigma\sigma'}\Gamma_{R}^{\sigma'}G_{a}^{\sigma'\sigma}]$,
where $\Gamma_{L(R)} =i[\sum_{L(R)}^{r}-\sum_{L(R)}^{a}]$,
$\sum_{L(R)}^{r}=(\sum_{L(R)}^{a})^{\ast}$ is the self-energy from
the left (right) lead,
$G_{r}^{\sigma\sigma'}(G_{a}^{\sigma'\sigma})$ is the retarded
(advanced) Green function of the whole system, and the lead effect
is incorporated into the self-energy of Green function
$G_{r}^{\sigma\sigma'}(G_{a}^{\sigma'\sigma})$ . The trace is over
the spatial degrees of freedom. The Green function above is computed
by the well-known recursive Green function method.$^{25,26}$ The
local density of electron states (LDOS) is described as$^{27}$
$\rho(\vec{r},E)=-\frac{1}{\pi}Im[G_r(\vec{r},\vec{r},E)]$, where
$E$ is the emitting energy of electrons. In our following
calculation, the $z$ axis is chosen as the spin-quantized axis, all
the energy is normalized by the hoping energy $t(t=1)$, and the
structural parameters of the wire are fixed at $L_1=L_3=10~a$,
$L_2=30~a$, $W_1=20~a$ and $W_2=9~a$ except the variables in Fig.
(5). The charge conductance and the vertical spin polarization are
defined as
$G^{e}=G^{\uparrow\uparrow}+G^{\downarrow\uparrow}+G^{\downarrow\downarrow}+G^{\uparrow\downarrow}$
and
$P_{z}=((G^{\uparrow\uparrow}+G^{\uparrow\downarrow})-(G^{\downarrow\downarrow}+G^{\downarrow\uparrow}))/G^{e}$,
respectively.

\section{RESULTS AND DISCUSSION}

The energy-subband dispersions of the narrow and wide part of the
quantum wire are plotted in Fig. 1(a) and Fig. 1(b), respectively.
The Rashba SOI strength $t_{so}=0.153$. It is shown that a linearly
Rashba spin-split subband is obtained in Fig. 1(a). However, the
spin-subband dispersion in Fig. 1(b) deviates considerably from the
typical linearly Rashba splitting. This is because the SOI strength
in the narrow region lies in the weak-coupling regime while it lies
in the strong-coupling regime in the wide regions due to their
different widths.$^{8}$

Figure 2 shows the calculated spin polarization as a function of the
emitting energy $E$ and the strength of Rashba SOI $t_{so}$. When
the emitting energy $E>0.09$, a vertical spin-polarized current can
be obtained in the right lead since the injected current from the
left lead is unpolarized. This vertical spin-polarized current
arising from the Rashba intermixing between different pairs of
transverse modes and the mirror symmetry with respect to the y-axis
is broken.$^{11}$ According to the left-right and time-reversal
symmetries, the relation
$G^{\downarrow\uparrow}=G^{\uparrow\downarrow}$ is obtained while
the relation $G^{\uparrow\uparrow}= G^{\downarrow\downarrow}$ cannot
be guaranteed, which can be proved by checking the numeric
calculation. Surprisingly, there appears a very large vertical
spin-polarized current when the emitting energy $E$ be about to 0.2.
In order to clarify this effect, the charge conductance as a
function of the emitting energy at $t_{so}=0.153$ is illustrated in
Fig. 3(a) and the corresponding spin polarization is shown in the
inset of Fig. 3(a). When $E<0.07$, the conductance is zero since all
the subbands of the narrow region are evanescent modes. When
$0.07<E<0.09$, only the lowest one pair of subbands of the wide and
narrow regions are propagating modes, so there is no spin polarized
current.$^{28}$ Moreover, when $E>0.09$, the lowest two pairs of
subbands of the wide regions become propagating modes and the
subband intermixing induced by the Rashba SOI arises, resulting in
the nonzero spin polarized current. However, there is still a pair
of propagating modes in the narrow region, which determines the
charge conductance of the whole system. Consequently, the vertical
spin polarization may be larger than 0.5. It is worth to note that a
`valley' like structure appears in the charge conductance when the
emitting energy just near the threshold of the third pair of
propagating modes in the wide regions, i.e., $E=0.209$. This effect
can be attributed to the constriction in the narrow part, leading to
the formation of a bound state in the right wide region. And the
bound state couples to the conductance ones through Rashba
intersubband mixing, giving rise to a structure-induced Fano
resonance. Amazingly, a very large vertical spin polarization
$|P_{z}|=0.77$ can be achieved in the right lead at this Fano
resonance. Fig. 3(b) displays the LDOS at $E=0.209$ and
$t_{so}=0.153$. It is obvious that electrons are confined in the top
of the right wide region under the present parameters.  Of course,
there are other structure-induced or SOI-induced Fano resonances
near the thresholds of the second and the fourth pair of propagating
modes in the wide region, i.e., $E=0.10$ and $E=0.36$, but the spin
polarization at this two points is much smaller than that at
$E=0.209$.

The spin polarization as a function of the strength of Rashba SOI at
$E=0.209$ is shown in Fig. 4. The spin polarization becomes larger
with the increasing of the strength of Rashba SOI and reaches its
maximum value $|P_z|=0.77$ as $t_{so}=0.153$, which consists with
the maximum value of the inset in Fig. 3(a). However, the spin
polarization becomes smaller as the Rashba SOI strength further
increases. It indicates that the spin polarization at the
structure-induced Fano resonance can be manipulated by variation of
the Rashba SOI strength. Apart from the strength of Rashba SOI, the
structural parameters of the investigated system can also be used to
control the magnitude of the spin polarization. Figure 5(a) shows
the calculated spin polarization as a function of the width of the
narrow region when $E=0.209$ and $t_{so}=0.153$. The spin
polarization becomes larger as the width of the narrow region
increases and reaches its maximum value at $W_2=9~a$. In addition,
there appears a sharp step as the width of the narrow region reaches
$12~a$. Furthermore, the spin polarization reaches zero when
$W_2=20~a$, namely, the spin-polarized current disappears, which
results from the structural and time-reversal symmetries. According
to Fig. 3(b), the bound state exists in the top of the right wide
region. Therefore, the spin polarization is only related to the
length of this region while has nothing to do with the length of the
left wide region. The calculated spin polarization as a function of
the length of the right wide region is plotted in Fig. 5(b). The
emitting energy and SOI strength are the same as that in Fig. 5(a).
The most interesting features are that a dip emerges in the spin
polarization when $L_3=10~a$ and the spin polarization oscillates
around zero when $L_3>14~a$.

The above calculation assumes perfectly clean system, where there is
no elastic or inelastic scattering. However, in a realistic system,
there will be many impurities in the sample. Consequently, the
effect of disorder should be considered in practical application.
The disorder could be incorporated by the fluctuation of the on-site
energies, which distributes randomly within the range width $w$
[$\varepsilon_{lm\sigma}= \varepsilon_{lm\sigma}+w_{lm}$ with
$-w/2<w_{lm}<w/2$]. Figure 6 shows the calculated spin polarization
as a function of the emitting energy for (weak and strong) different
disorders $w$. The SOI strength is set as $t_{so}=0.153$. It can be
seen that the spin polarization $|P_z|$ at the structure-induced
Fano-resonances decreases with the increasing of disorder. However,
the spin polarization at the second Fano-resonance almost equals to
$0.5$ even when $w=0.6$, which implies that the spin-polarized
current can still survive in the presence of strong disorder.

\section{CONCLUSION}
In conclusion, we have studied the spin-polarized electron transport
through a WNW quantum wire under the modulation of Rashba SOI. The
presence of the SOI and the asymmetry in the longitudinal direction
leading to a vertical spin-polarized current in the right lead. In
particular, we have shown that a very large vertical spin
polarization can be obtained at the structure-induced resonance and
its degree can be controlled by both the strength of Rashba SOI and
the structural parameters even in the presence of strong disorder.
This effect may provide an efficient method to design a spin filter
without applying an external magnetic field and without attaching
ferromagnetic contacts.

\begin{flushleft}
\section*{ACKNOWLEDGMENT}
\end{flushleft}

This work was supported by the National Natural Science Foundation
of China under Grant No. 10774112.

\newpage

\begin{figure*}
\includegraphics[width=3.5in]{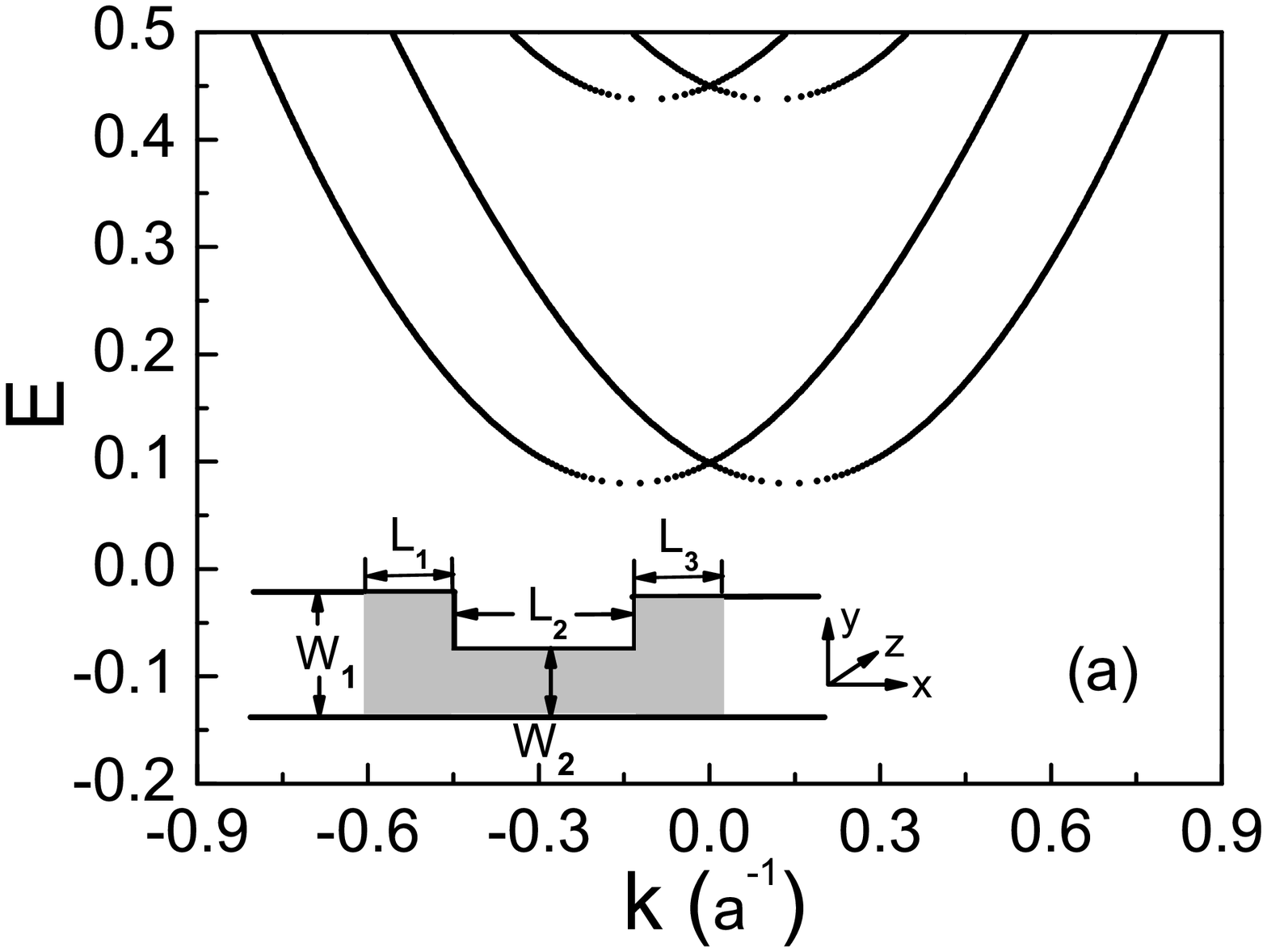}
\includegraphics[width=3.5in]{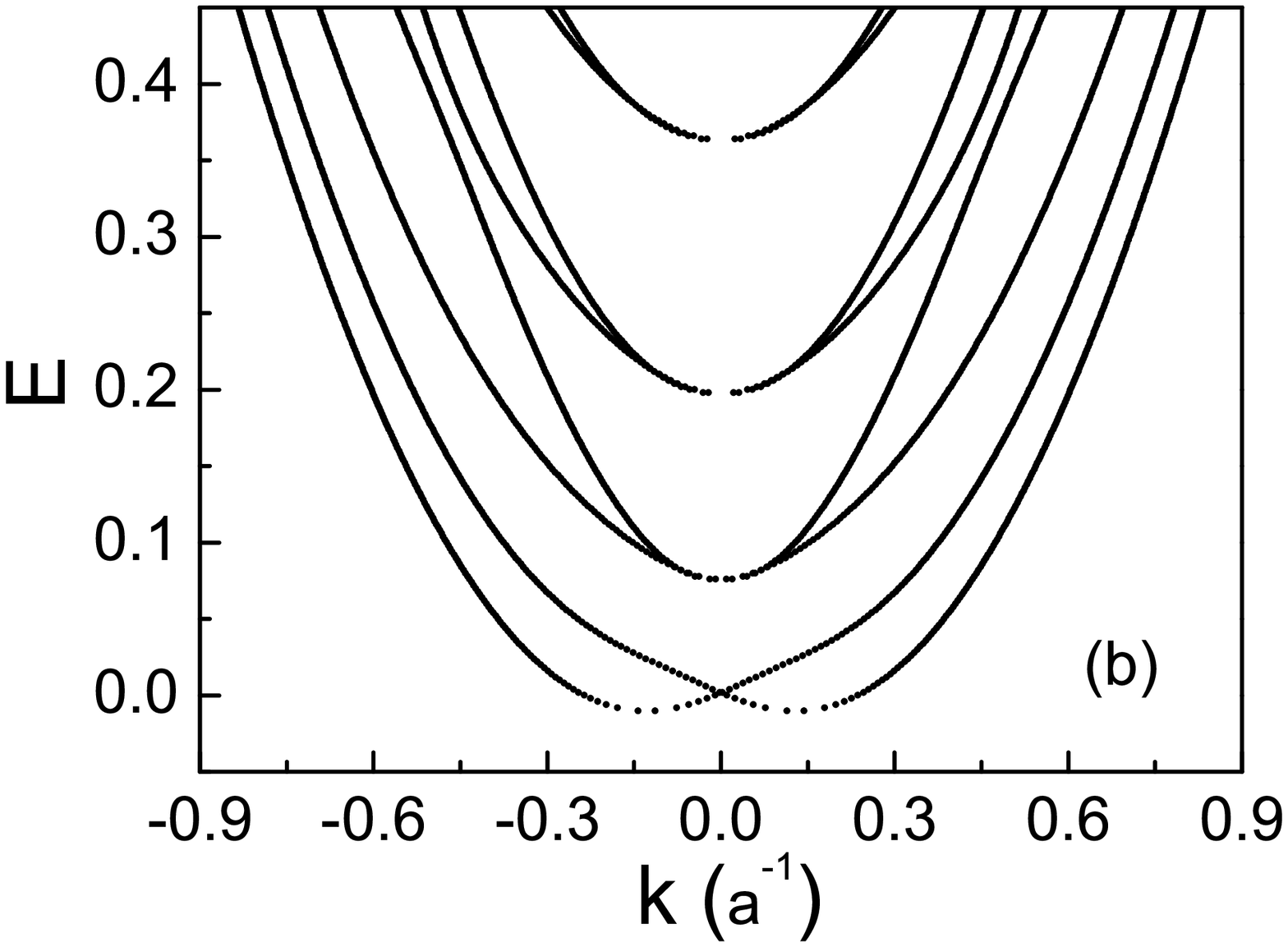}
\caption{\label{fig:wide}(a) Energy-subband dispersion for the
narrow part of the quantum wire. Inset: schematic view of the system
considered here. (b) Energy-subband dispersion for the wide part of
the quantum wire. The Rashba SOI strength $t_{so}=0.153$.}
\end{figure*}

\begin{figure*}
\includegraphics[width=4in]{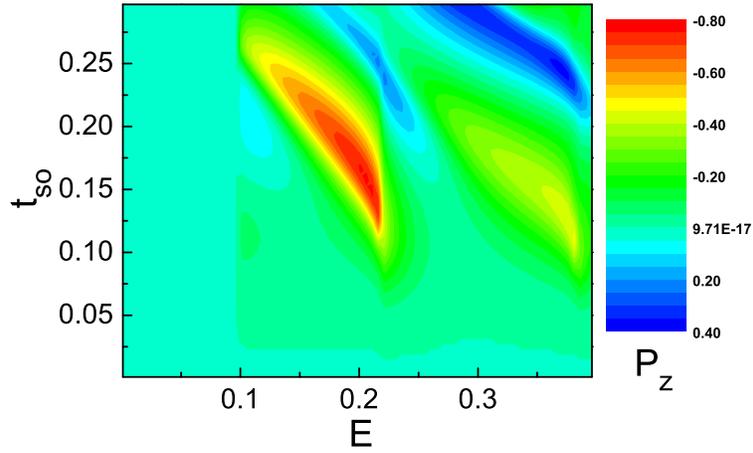}
\caption{\label{fig:wide}(Color online) The calculated spin
polarization as a function of the emitting energy and Rashba SOI
strength for spin-unpolarized electron injection.}
\end{figure*}

\begin{figure*}
\includegraphics[width=3.5in]{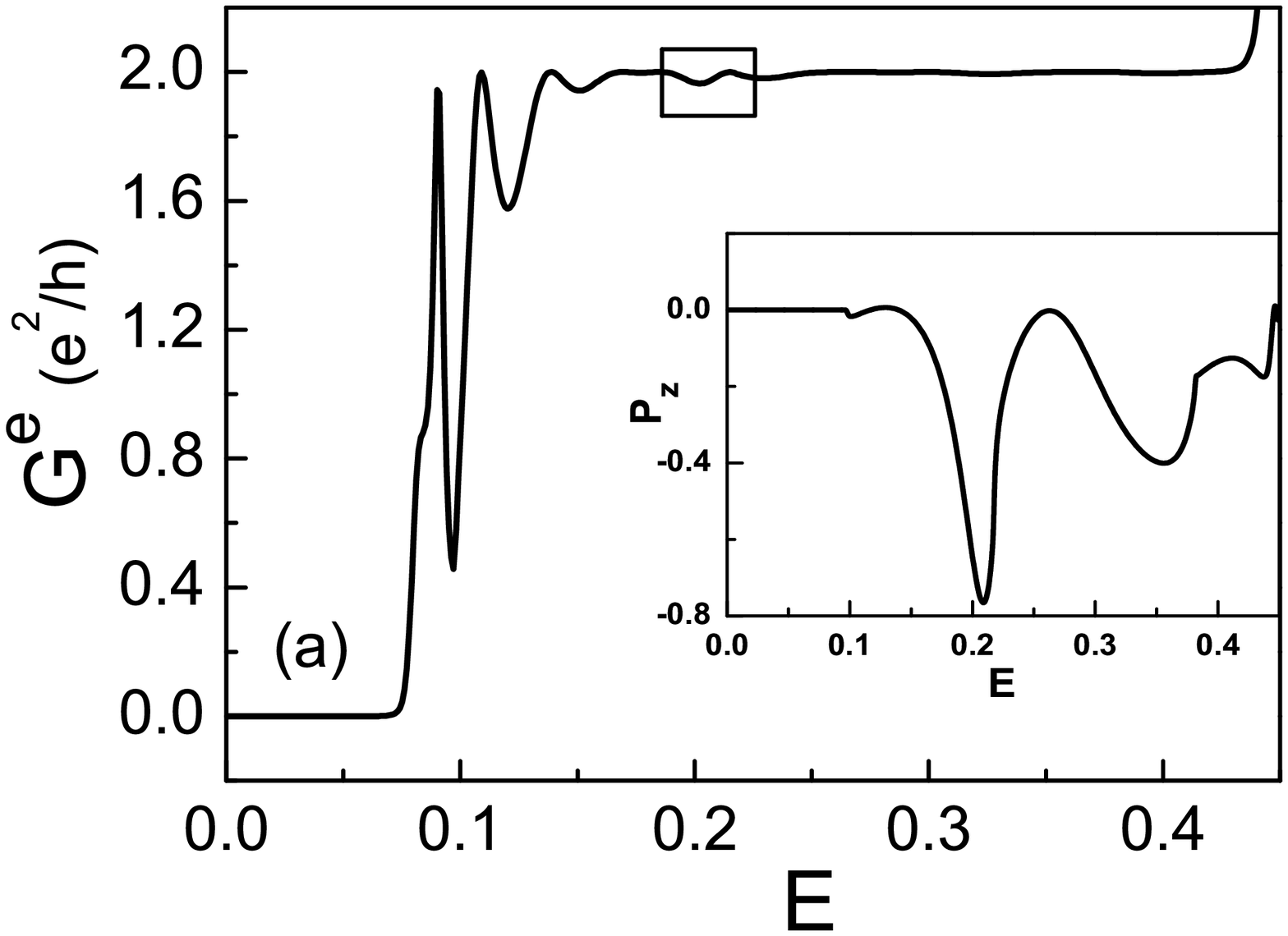}
\includegraphics[width=3.5in]{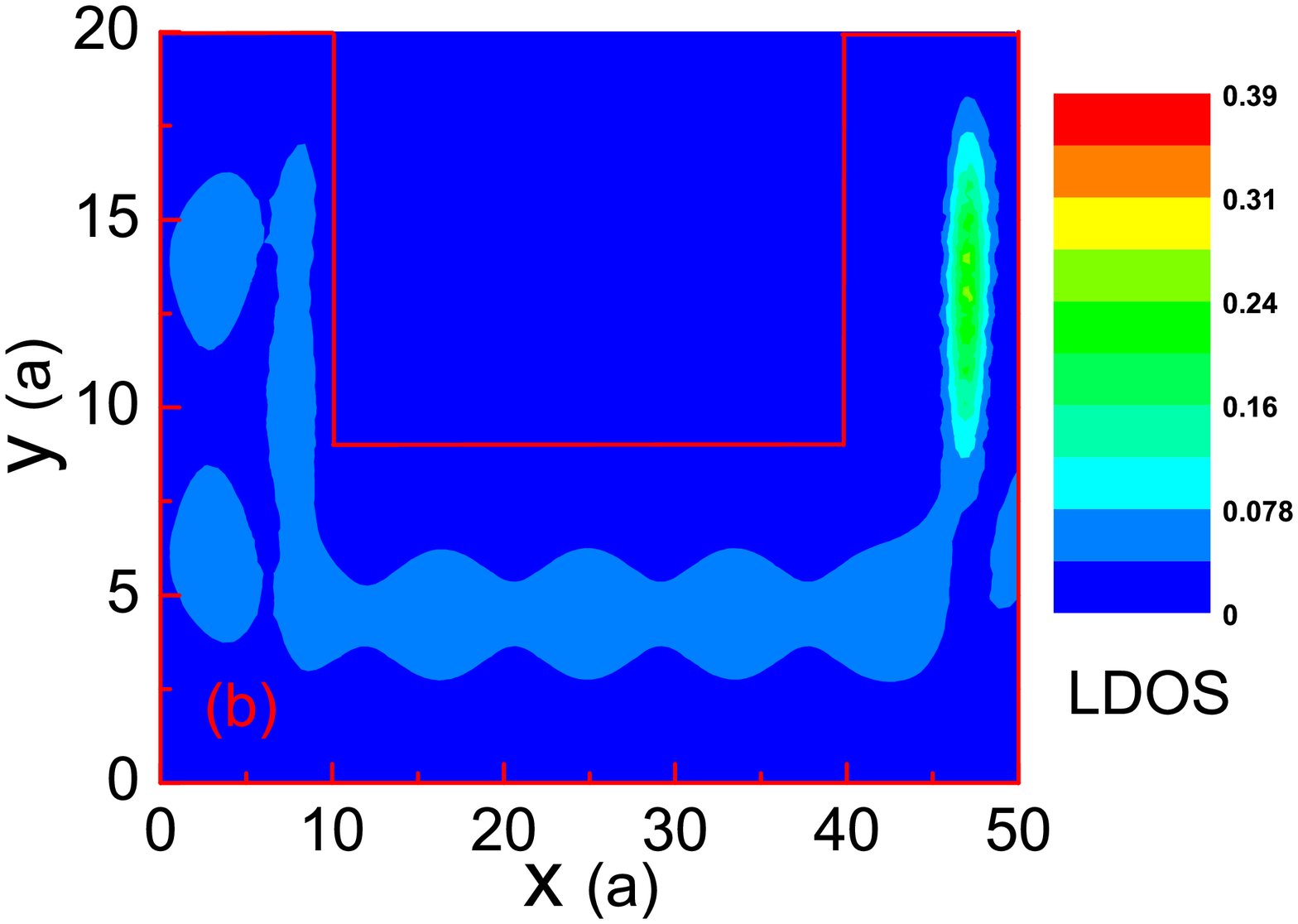}
\caption{\label{fig:wide}(a) The calculated charge conductance as a
function of the emitting energy. Inset: the corresponding spin
polarization as a function of the emitting energy. The Rashba SOI
strength $t_{so}=0.153$. (b) (Color online) The local density of
electron states. The emitting energy $E=0.209$ and the Rashba SOI
strength $t_{so}=0.153$.}
\end{figure*}

\begin{figure*}
\includegraphics[width=4in]{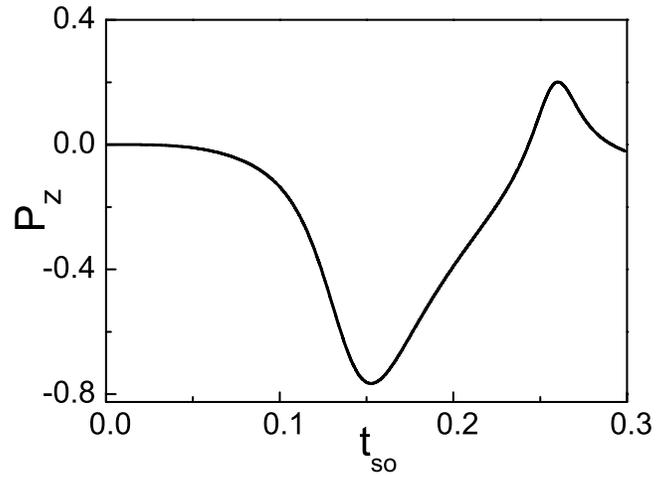}
\caption{\label{fig:wide}The calculated spin polarization as a
function of the strength of Rashba SOI. The emitting energy
$E=0.209$.}
\end{figure*}

\begin{figure*}
\includegraphics[width=3.5in]{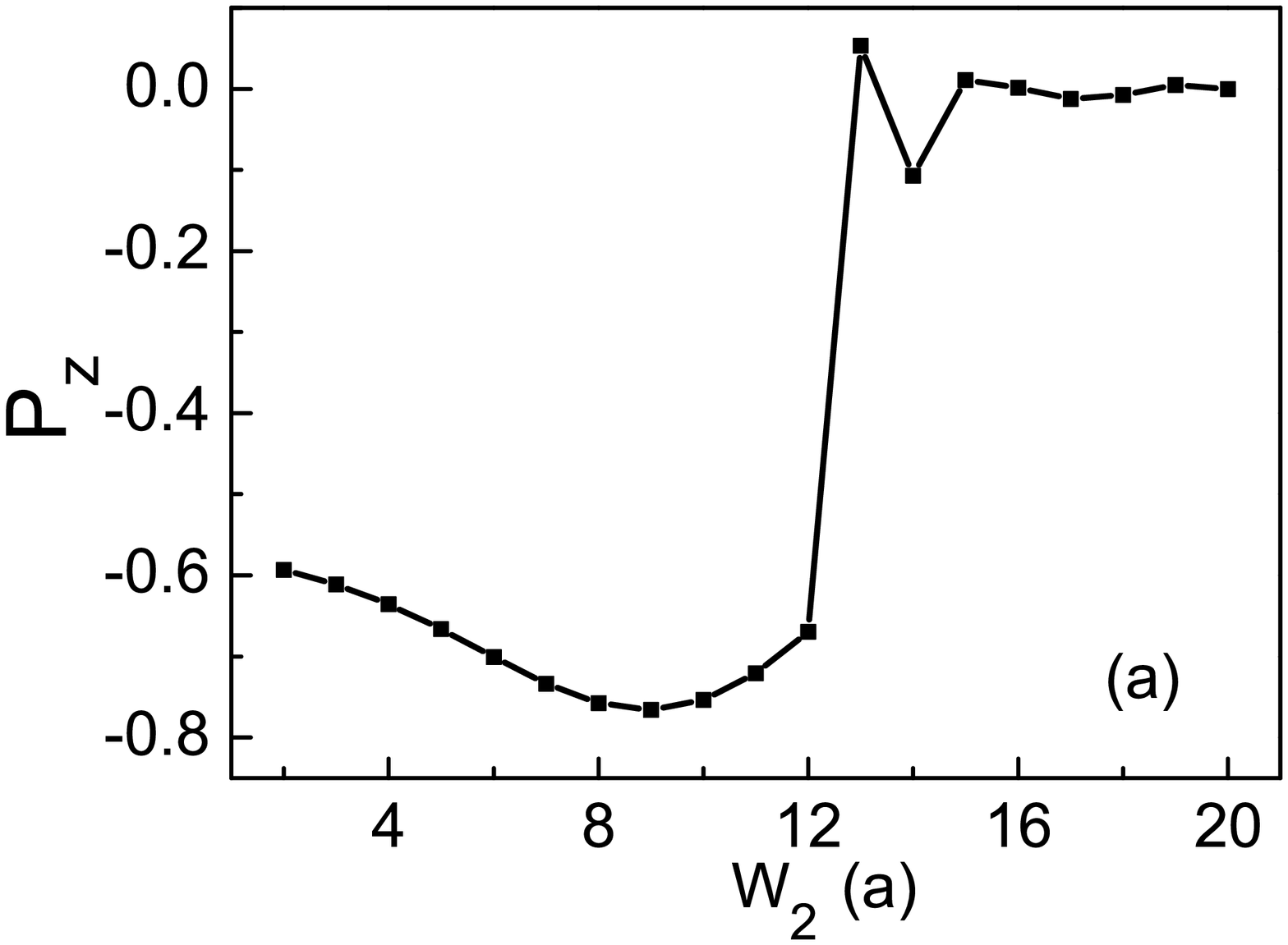}
\includegraphics[width=3.5in]{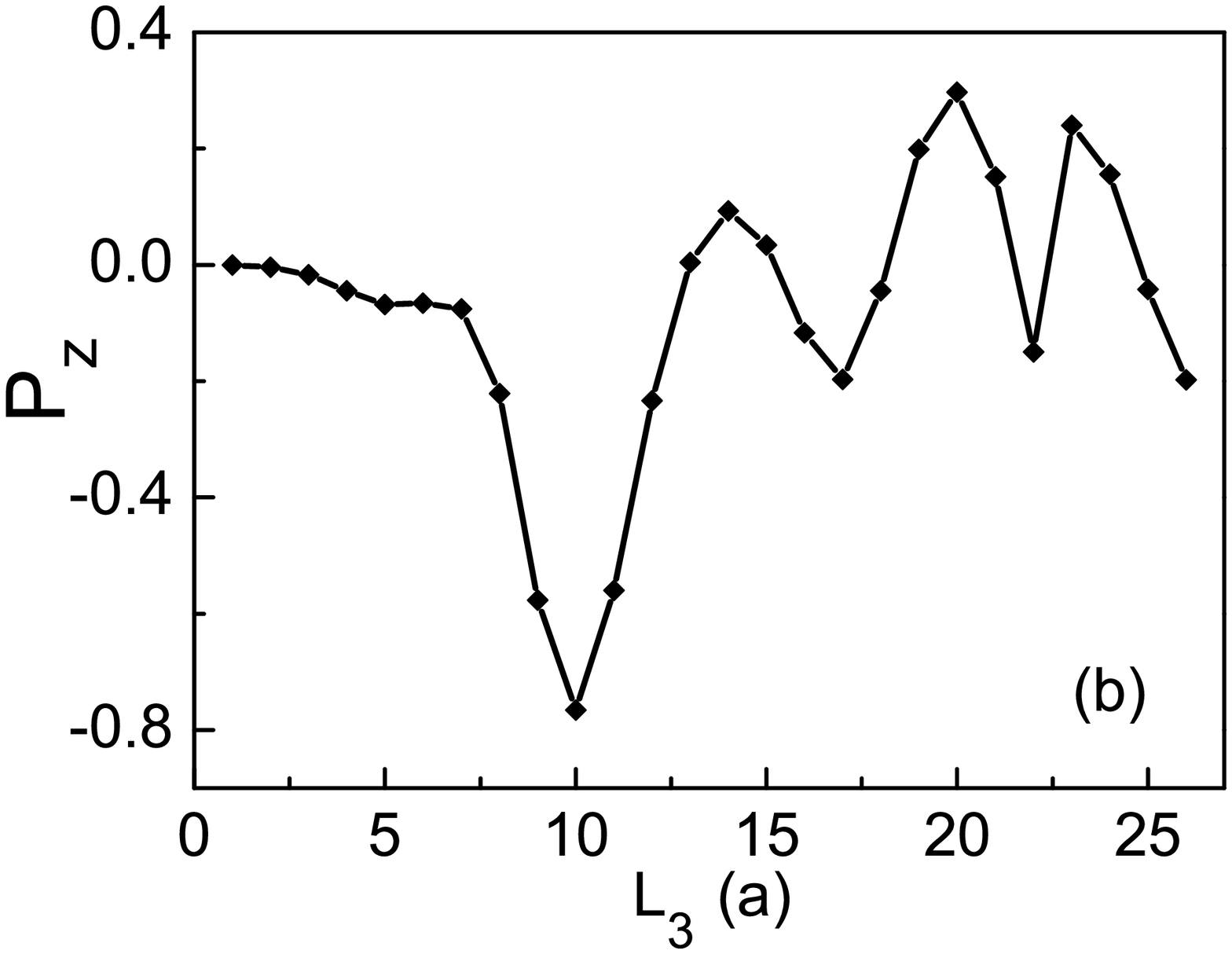}
\caption{\label{fig:wide}(a) The calculated spin polarization as a
function of the width of the narrow region. (b) The calculated spin
polarization as a function of the length of the right wide region.
The parameters are the same as that in Fig. 3(b).}
\end{figure*}

\begin{figure*}
\includegraphics[width=4in]{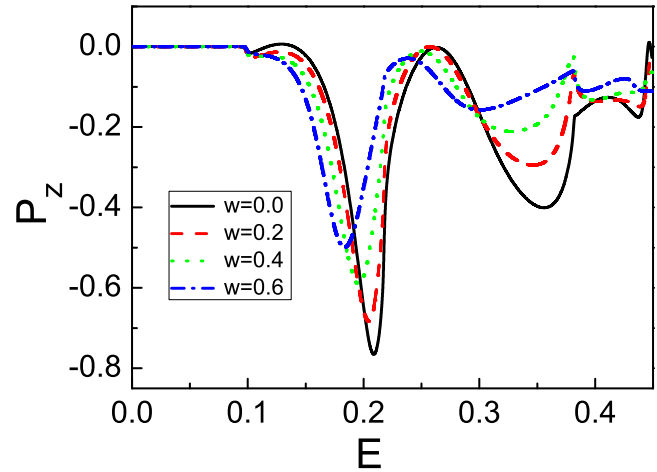}
\caption{\label{fig:wide}(Color online) The calculated spin
polarization as a function of the emitting energy for different
disorder strengths. The Rashba SOI strength $t_{so}=0.153$.}
\end{figure*}

\end{document}